\documentclass[12pt]{iopart}

\usepackage{iopams}
\usepackage{graphicx}

\begin{document}


\title[Anomalous Scaling Behavior in Polymer Thin Film Growth]
{Anomalous Scaling Behavior in Polymer Thin Film Growth by Vapor Deposition}

\author{Seung-Woo Son, 
Meesoon Ha\footnote{Author to whom any correspondence should be addressed.}, 
and Hawoong Jeong}
\address{Department of Physics, Institute for the BioCentury,
KAIST, Daejeon 305-701, Korea}
\ead{sonswoo@kaist.ac.kr, msha@kaist.ac.kr, and hjeong@kaist.ac.kr}
\date{\today}

\begin{abstract}
\noindent
As a first step to understand anomalous kinetic roughening
with multifractality in recent experiments of the
vapor deposition polymerization (VDP) growth,
we study a simple toy model of the VDP growth in a (1+1)-dimensional lattice,
along with monomer diffusion, polymer nucleation,
limited active end bonding, and shadowing effects.
Using extensive numerical simulations,
we observe that the global roughness exponent is different from the local one.
It is argued that such anomalies in VDP growth are attributed to
the instability induced by the nonlocal shadowing effects
on active ends of polymers. As varying the ratio of diffusion coefficient
to the deposition rate by cosine flux, we also discuss the role of diffusion
in kinetic roughening of the polymer thin film growth, which is quite different
from that of the metal or semiconductor film growth. Finally, we suggest
its (2+1)-dimensional version, which can be directly compared
with experimental results.
\end{abstract}

\\

\noindent\textbf{ArXiv}: 0811.2169 



\maketitle

\section{Introduction}
In the last few years,
kinetic roughening of non-equilibrium steady states
for the growth of thin films and multilayers has been
an issue of considerable interest~\cite{FnV1991B+BnS1995B,Meakin1998B}.
This is motivated by the demand for smooth or regularly structured surfaces
and interfaces for miniaturized functional films in science and technology.
Such interest is explained by the relevance to surface film characterization 
at the submicron level, and the mechanisms that determine the film morphology and
can contribute to achieving better control of the film properties
in real applications.

Although lots of theoretical and experimental studies
have shown the existence of kinetic roughening and in many cases revealed
the occurrence of scaling exponents corresponding to a few universality classes,
there is no general picture of kinetic roughening for the growth of polymer thin films.
This is because the major efforts have been focused on the growth of metal
and semiconductor thin films. Now that the polymer thin films are
growing technological interest, as regards molecular devices
and microelectronic interconnects~\cite{Wong1993B},
the few such studies are known as pioneering works where also kinetic roughenings 
with various scaling behaviors are shown~\cite{Collins1994+Biscarini1997,Zhao2000,IJLee2008}.
Among the many techniques for the polymer thin film growth, vapor deposition polymerization (VDP)
best describes the process of coating with poly ($p$-xylylene) (PPX),
also known by the trade name Parylene~\cite{VDP-PPX},
where the monomer from the gas phase condenses on the substrate,
reacts to form a high molecular weight as an oligomer, and becomes a part of the polymer.
In the present study, we mimic such VDP processes, in term of a modified MBE-type growth model,
discuss kinetic roughening of the polymer thin film growth by vapor deposition,
and give a guideline for the VDP growth model studies for explaining the experimental data
from the growth of polymer thin films.

This paper is organized as follows. In section~\ref{sec:model},
we describe our model for the VDP growth in a (1+1)-dimensional lattice
and show the evolution of surface morphologies with and without
shadowing effects caused by a cosine flux. In section~\ref{sec:results},
numerical results are presented for kinetic roughening with multifractality
as measuring surface roughness, height difference correlation functions, the density
profile, height and step distributions. Finally, we discuss
the physical origin of the anomalous scaling behaviors as well as
polymer characteristics, and suggest a possible extension of the VDP model
in a (2+1)-dimensional lattice, to be compared with recent experimental data of PPX-C film growth.
We conclude the paper in section~\ref{sec:conclusion} with a brief summary and remarks.

\section{Model}
\label{sec:model}

We mimic the polymer thin film growth by the VDP process
in terms of a simple toy model was proposed by Bowie and Zhao~\cite{Bowie2004},
for a (1+1)-dimensional lattice with $L$ sites, where we use a periodic boundary condition
in a spatial direction, $x$, and add the coalescence process of polymers
to the original model.

During the VDP process, the monomer transport in the vacuum is very similar to
the conventional physical vapor deposition (PVD) process,
i.e., molecular beam expitaxy (MBE) process for metals or semiconductors~\cite{FnV1991B+BnS1995B}.
However, they are quite large differences in the nucleation and growth processes
after the monomer is condensed on the substrate or the film surface. In the PVD/MBE process,
monomers are stable once they attach to the nearest neighbors of any nucleated sites,
so that the films get dense and compact as monomer diffusion increases. In contrast,
they become stable in the VDP process only when they reach one of two active ends of a polymer chain,
and the films get rough as monomer diffusion increases since it occurs along the polymer bodies.
Other surface dynamics can also affect the growth differently in the two cases.
While surface diffusion, edge diffusion, step barrier effect are relevant to the PVD/MBE case,
intermolecular interaction and chain relaxation are relevant to in the VDP case
besides monomer diffusion. Such differences give a distinct dynamic behavior
for the VDP film morphology.

\subsection*{Dynamic rules and updates}

\begin{figure*}[b]
\begin{center}
\includegraphics[width=0.8\columnwidth]{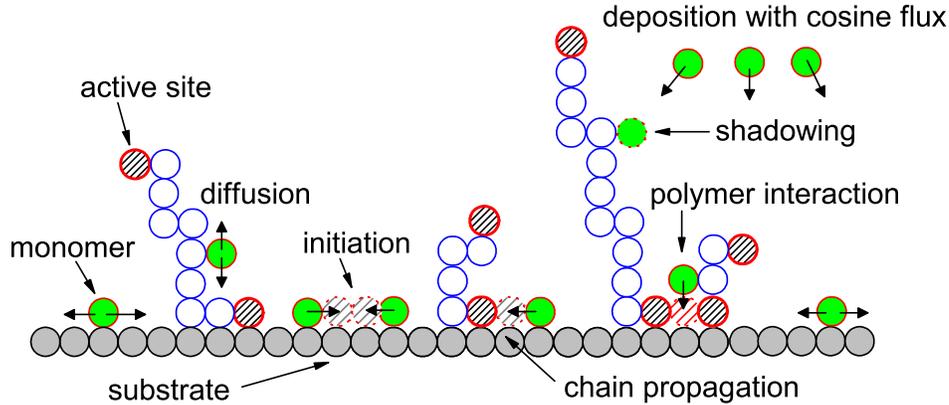}
\end{center}
\caption{Five dynamic rules are illustrated as
solid circles for monomers, open circles with thin lines for polymer bodies,
and patterned circles with thick lines for active ends. \label{fig:model}}
\end{figure*}

For simplicity, we omit the chain relaxation in our model
and consider only the following five processes (see figure~\ref{fig:model}):
\begin{description}
\item {{\bf Deposition}.
At each step, a monomer is activated into the system with an angle of
incidence $\theta$ to the vertical direction, which follows the distribution of $cos(\theta)$, 
not a collimated flux. This incidence of monomers with angle distribution is called
as a {\em cosine flux}~\cite{cosineF} with the deposition rate $F$,
the number of incident monomers per site for unit time.}
\item {{\bf Surface Diffusion}.
Before the activated monomers are stabilized,
an incident monomer deposited onto the polymer body sides or substrate
randomly wanders from one site to another site
along the polymer bodies or substrate with diffusion coefficient $D$
at each deposition step,
where $D$ is the number of hops per monomer for unit time.
The surface growth is controlled by the ratio of the diffusion coefficient
to the deposition flux, $G=D/F$. From now on we set $F=1$ for convenience,
such that $G=D$.}
\item {{\bf Nucleation}.
When two monomers are met on substrate or polymer bodies,
they form a dimer as a polymer seed, i.e., oligomer, which is called as
{\em nucleation (initiation)}. In contrast to the MBE growth where
atoms can attach to the nearest neighbors of the nucleated sites,
in the VDP growth the stabilization reaction occurs only at the active ends of
a polymer chain, so-called active sites. Such active bonding in the VDP growth
is a key ingredient as well as the cosine flux for monomer deposition.}
\item {{\bf Propagation}.
When a monomer reaches one of the active ends of a polymer,
it is stabilized as part of the polymer and at the same time
it becomes the active end of the polymer.
This is called as {\em chain propagation}.}
\item{{\bf Coalescence}.
In the process of the chain propagation,
it is possible that an active end of polymer meets
that of another polymer. Then two polymers are merged into
one long polymer. This process is called as
{\em coalescence (polymer interaction)}.
It is worthy of note here that, for linear polymers,
only the two ends of the chain are active, and
are ready for reacting with monomers or other polymers.
However, we do not allow the polymer loop. In other words,
if one active end of a polymer meets
the other side active end of itself, the two active ends
cannot merge into a stabilized polymer loop and
such a try is rejected.}
\end{description}

Performing Monte Carlo (MC) simulations for the VDP growth model,
we use the random sequential (continuous time) updating method,
in terms of the deposition probability of an incident monomer, $P_F$,
and the diffusion probability of an ad-monomer, $P_D$,
with the definitions $P_F=\frac{FL}{DN_m+FL}$ and $P_D=1-P_F=\frac{DN_m}{DN_m+FL}$,
respectively. Here $N_m$ is the number of ad-mononers and $L$ is the system size,
We rewrite the probabilities by the ratio $G$ of the diffusion coefficient $D$
to the deposition rate $F$, $G=D/F$, and the ad-monomer density $\rho_m=N_m/L$:
$$
P_F=\frac{1}{G\rho_m+1},~{\rm and}~P_D=\frac{G\rho_m}{G\rho_m+1}.
$$

The detailed procedure of our MC simulations is as follows.
First, generate a random number, $p\in (0,1]$. If $p<P_F$,
a monomer is deposited on the polymer bodies or substrate
from the cosine flux with a randomly chosen angle.
Otherwise, an ad-monomer randomly chosen from $N_m$
monomers diffuses in a randomly chosen direction.
Then, the final surface configuration is governed by the above five VDP processes.
The MC time is updated as the unit of a monolayer (ML)
after every $L$ th monomer is deposited.

\begin{figure*}[t]
\begin{center}
\includegraphics[width=0.75\columnwidth]{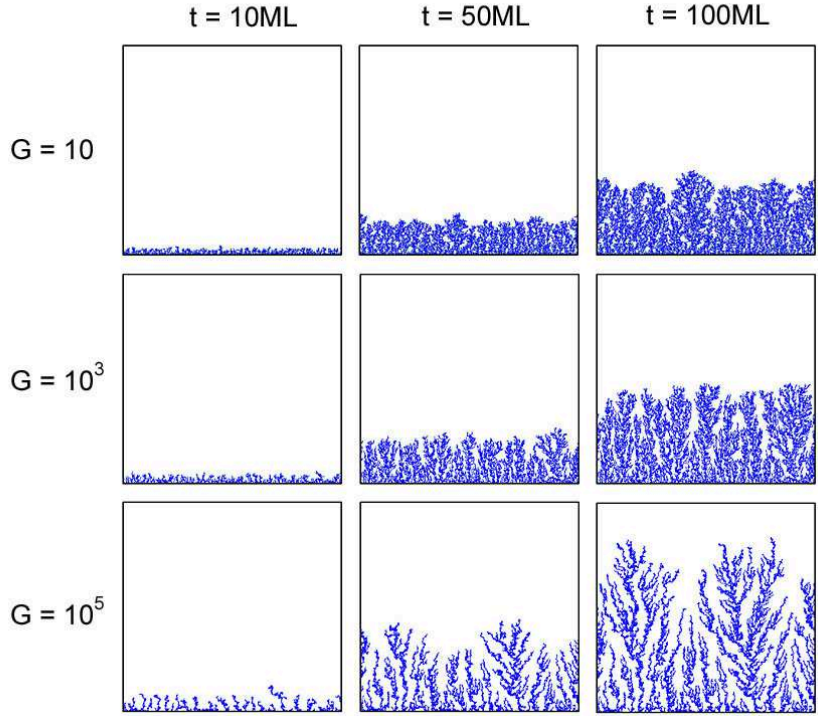}
\end{center}
\caption{Snapshots of the VDP growth for $L=512$
at three specific times, $t=10,~ 50,$ and $100$ ML
for three values of $G=D/F$ with $F=1$.
From top to bottom panels, $G=10,~10^3$, and $10^5$.
\label{snapshots}}
\end{figure*}
\begin{figure*}[]
\begin{center}
\includegraphics[width=0.8\columnwidth]{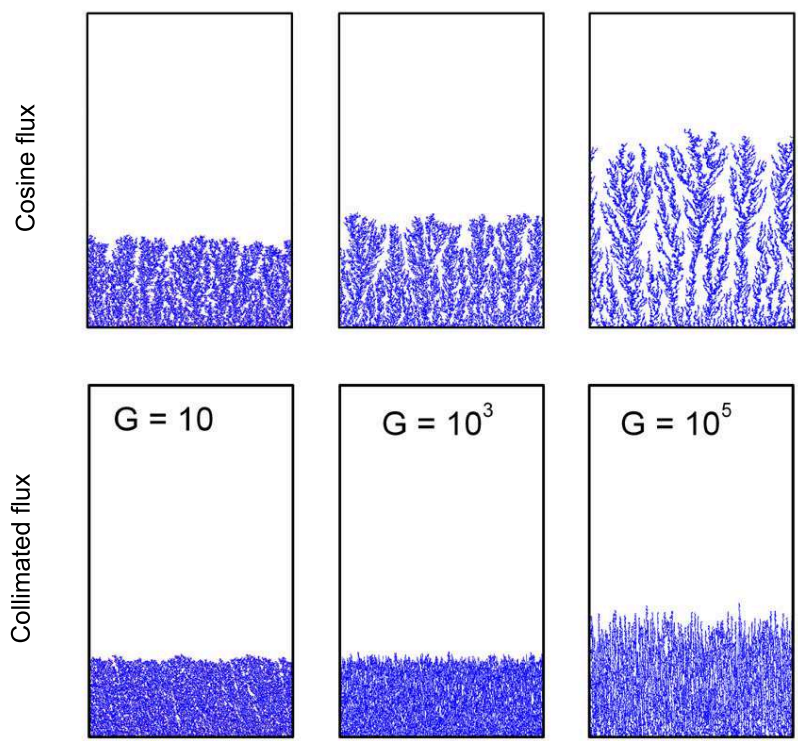}
\end{center}
\caption{Effect of the flux angle of incidence on the VDP growth model.
While the VDP growth with the cosine flux shows
the characteristic columnar structures (top panels),
such structures disappear when the angle of incidence is set to zero, i.e.,
for vertically collimated flux (bottom panels).
From left to right, $G=10, 10^3$, and $10^5$ for $L=512$
at $t=180$ ML.
\label{flux}}
\end{figure*}
\begin{figure*}[]
\begin{center}
\includegraphics[width=0.8\columnwidth]{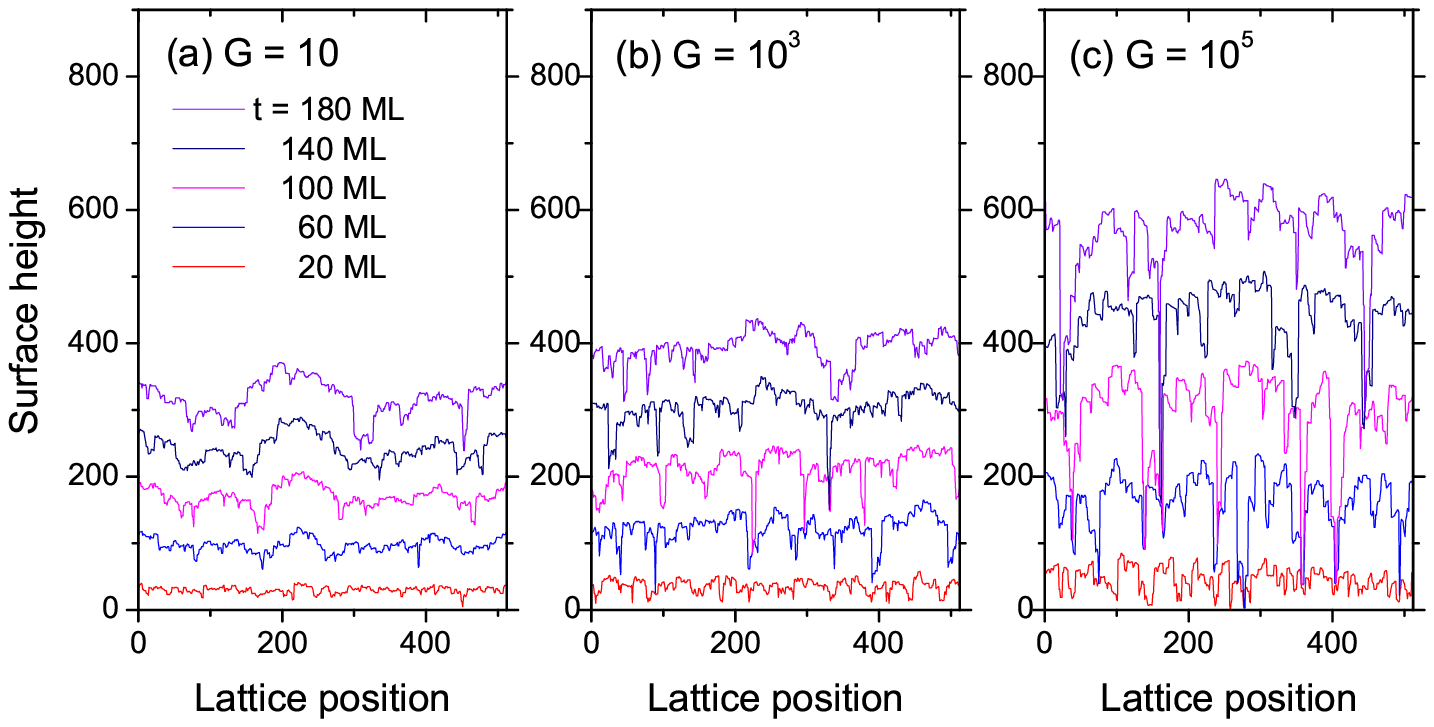}
\end{center}
\caption{Evolution of surfaces in the VDP growth model.
For $L=512$, (a) $G=10$, (b) $G=10^3$, and (c) $G=10^5$
at $t=20,~60,~100,~140$ and $180$ ML from bottom to top, respectively.
\label{evolution}}
\end{figure*}

\subsection*{Surface morphology}

Before staring the detailed analysis and the main discussion,
we check how the VDP growing surface evolves.
In plotting the snapshots of the VDP model growth
in figure~\ref{snapshots} for various $G$ values
at three different stages of the film growth,
we observe that the films exhibit tree-like characteristic morphologies
and columnar structures with many voids and overhangs
for all three cases of $G$ as time elapses.
Moreover, as $G$ (diffusion coefficient) increases,
the surface height grows rapidly and the columnar morphology
becomes rougher and less dense. In order to work out
the origin of the characteristic columnar structure,
we investigate the effect of the flux incident angle distribution
on the VDP growth. When we fix the monomer incident angle in a single
vertical direction such as that of a collimated flux,
the surface columnar structures disappear,
as shown in figure~\ref{flux}, for all three cases of $G$.
The evolution of surfaces by the VDP model growth is shown
in figure~\ref{evolution}, where we assume that the surface height is the
single value of the highest position at the lateral site.
One can see that, as $G$ increases and $t$ elapses,
the columnar and grooved structure becomes much clearer.

In section~\ref{sec:results}, we analyze this unusual VDP growing surface quantitatively
with conventional physical quantities in surface growth models as well as
polymer properties.

\section{Numerical Results}
\label{sec:results}

We perform numerical simulations of with various system sizes
up to $L=1024$ for three values of $G$, where numerical data are averaged over 100 samples.
Unlike the PVD/MBE growth model case, the VDP growth model case requires
active end site tracking and polymer indexing, so the largest system size
in our MC simulations becomes much smaller than that
in ordinary surface growth models.

\subsection{Surface roughness and height-difference correlation function}

\begin{figure*}[b]
\begin{center}
\includegraphics[width=0.37\columnwidth]{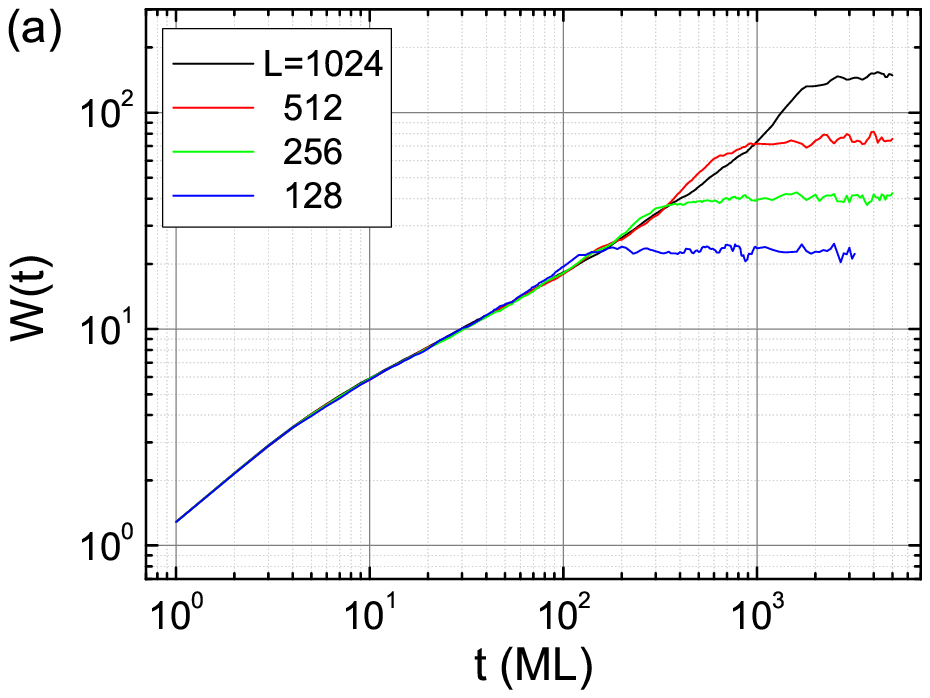}~~~~~
\includegraphics[width=0.38\columnwidth]{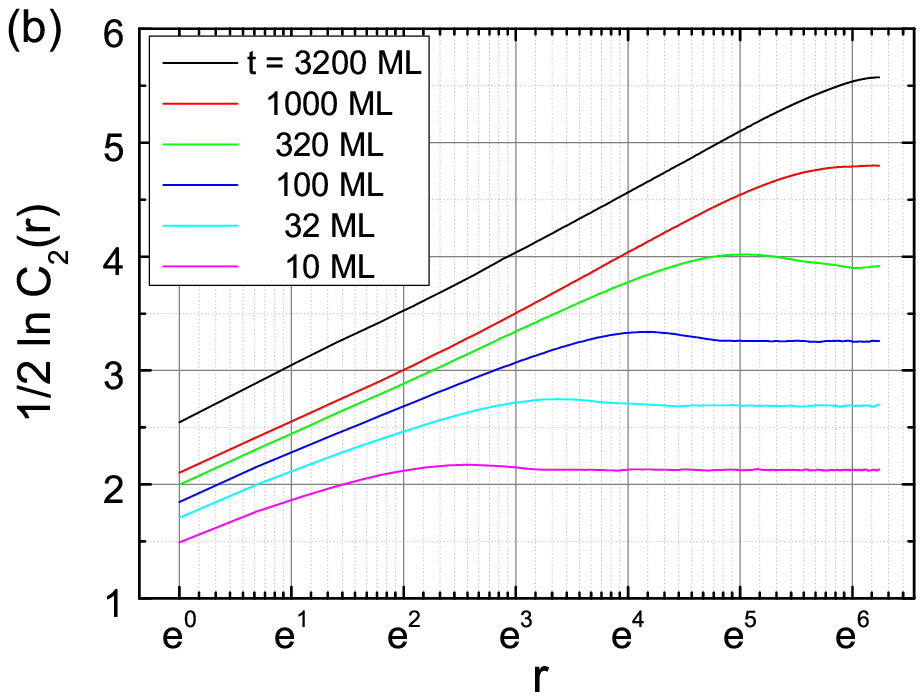}
\end{center}
\caption{Double-logarithmic plots of $W$ and $C_2$ for the case of $G=10$.
(a) surface roughness versus time for $L=128,~256,~512$, and $1024$ from bottom to top,
and (b) height difference correlation function versus $r$ for $L=1024$. \label{G10WtnC2r}}
\end{figure*}

We first measure the surface roughness (width) defined as
$$W^2(t) \equiv \langle{\overline{[h(x,t)-\bar{h}(t)]^2}}\rangle,$$
where $\bar{f}$ is the spatial average, i.e.,
$\bar{f}=\frac{1}{L}\sum_x f(x)$, and $\langle ...\rangle$
represents the statistical sample average. The width $W(t)$
in the VDP growth for $G=10$ plotted in figure~\ref{G10WtnC2r} (a),
which shows clearly three regimes as $L$ increases: the initial
growth, the VDP growth, and the saturation. Unlike the conventional surface growth,
the VDP growth exhibits anomalous dynamic scaling,
where the VDP growth regime appears after about five monolayers (ML),
irrespectively of the system sizes, and it undergoes some unusual behavior
before $W(t)$ saturates to $W_{\rm sat}$ due to the finite-size effect.
The global dynamic scaling of the VDP surface roughness is governed
by the global roughness exponent $\alpha_{\rm global}$,
from the system size dependence of the saturated width ($W_{\rm sat}\sim L^{\alpha_{\rm global}}$)
and the global dynamic exponent $z_{\rm global}$ from the system size dependence of
the saturation time ($t_{\rm sat}\sim L^{z_{\rm global}}$).

In order to investigate the local dynamic scaling of the VDP growth,
we also measure the two-point height difference correlation function
defined as
$$C_2(r,t)=\langle {\overline{|h(x+r,t)-h(x,t)|^2}} \rangle,$$
which follows $C_2 (r,t) \sim r^{2 \alpha_{\rm local}}$ for $ r < \xi(t)$
and $C_2(r,t) = 2 W^2(t)$ for $ r > \xi(t)$. Here $\xi(t)$ is the correlation length,
scaling as $\xi(t)\sim t^{1/z_{\rm local}}$. Figure~\ref{G10WtnC2r} (b) shows
how height correlations and the correlation length are developed at various times
for $G=10$.
\begin{figure*}[]
\begin{center}
\includegraphics[width=0.315\columnwidth]{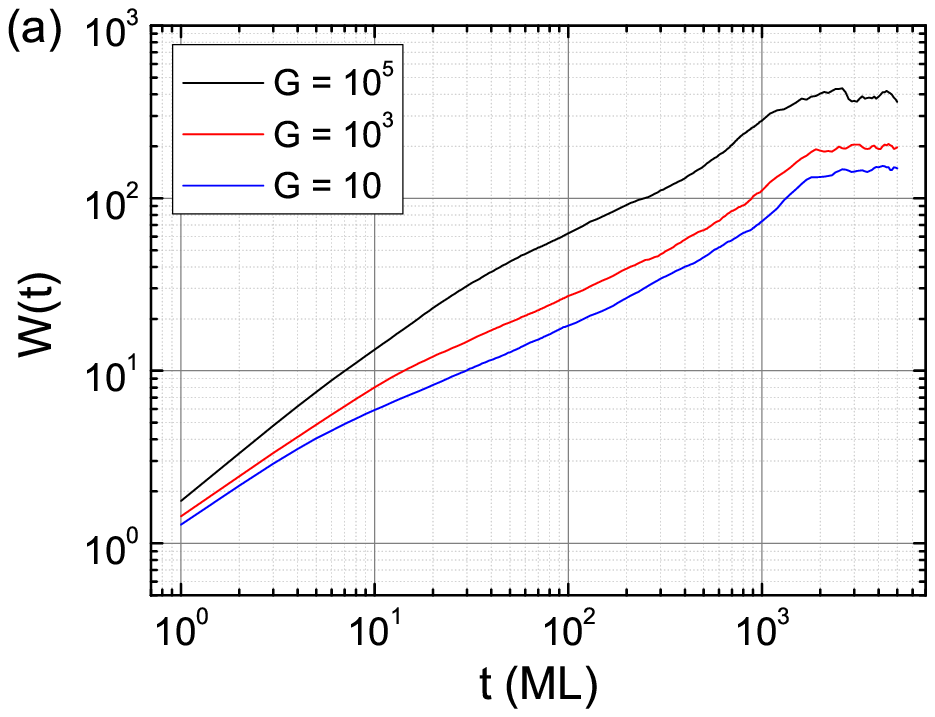}\hfill
\includegraphics[width=0.315\columnwidth]{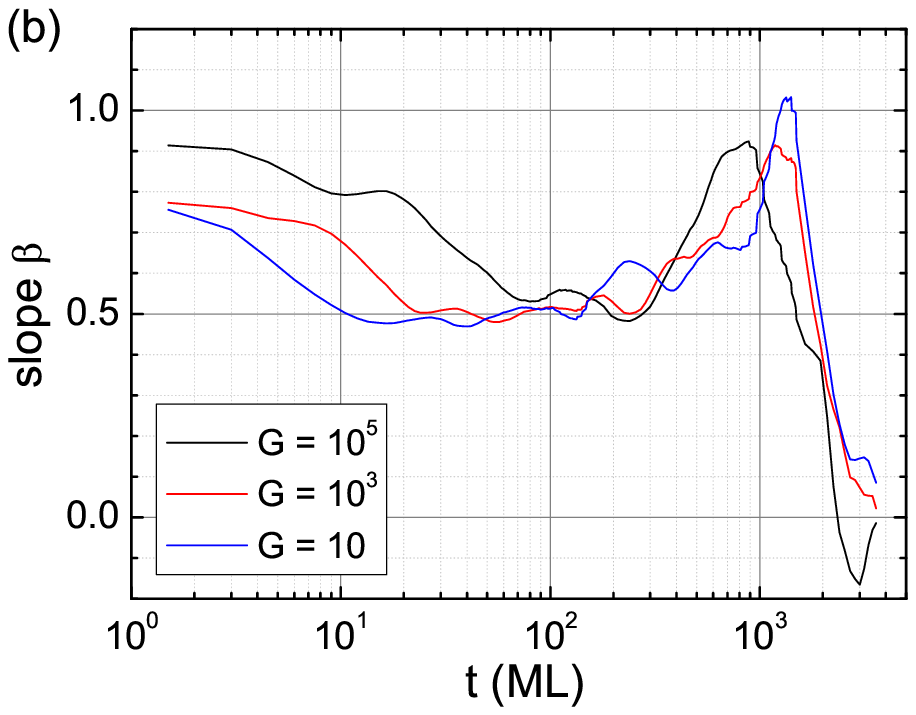}\hfill
\includegraphics[width=0.33\columnwidth]{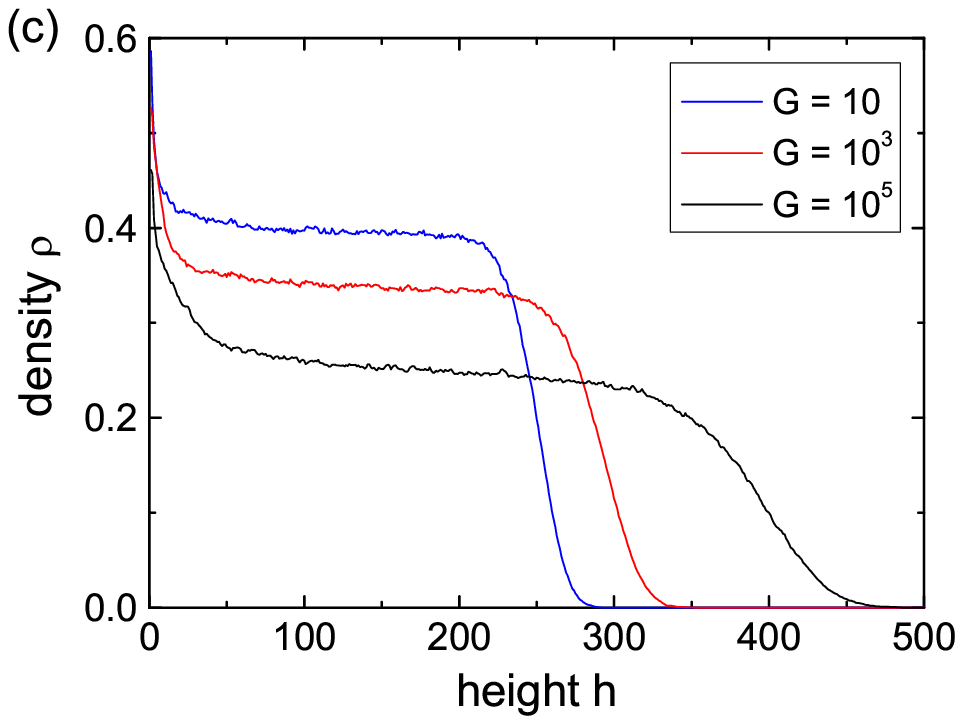}
\end{center}
\caption{For $L=1024$,
(a) double-logarithmic plots of $W$ against $t$,
(b) semi-logarithmic of the effective growth exponent $\beta$ against $t$,
(from bottom to top, $G=10,~10^3,$ and $10^5$),
and (c) the density profile at $t=100$ ML against surface height $h$
(from top to bottom, $G=10,~10^3,$ and $10^5$).
\label{variousG}}
\end{figure*}
\begin{figure}[b]
\begin{center}
\includegraphics[width=0.4\columnwidth]{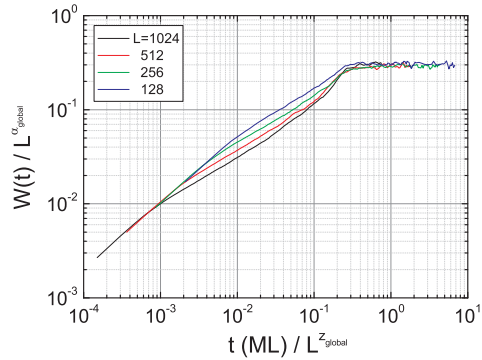}
\end{center}
\caption{Data collapse of surface roughness:
double-logarithmic plots of $WL^{\alpha_{\rm global}}$ versus $t/L^{z_{\rm global}}$
with $\alpha_{\rm global}=0.89$ and $z_{\rm global}=1.27$ for $G=10$.
\label{G10W-collapse}}
\end{figure}
For three values of $G$, the global scaling behavior in the VDP growth
is compared with the local one. Figure~\ref{variousG} shows
clearly that as $G$ increases, the surface becomes rough much faster
with the larger value of $W$, and less dense at each level of the surface height.
Moreover, from figure~\ref{variousG} (b) and (c), we observe
that the initial growth regime gets extended as $G$ increases,
while at the real scaling regime by the VDP growth, the effective
growth exponent $\beta$ becomes all the same as $\beta\simeq 0.5
(\ne\alpha_{\rm global}/z_{\rm global})$, irrespectively of the value of $G$.
This implies that at the early stage of the growth, the shadowing effect by the cosine flux
is negligible since there are not many polymers, but later on, the shadowing effect governs
the surface growth as well as the active bonding, once polymers form.
In the VDP growth regime, the density profile at each height level shows
the difference of dynamic process like a stratum reflects
the historical event (see figure~\ref{variousG} (c)). Until a polymer forms,
the effect of the cosine flux is negligible and the monomer diffusion is dominant,
which explains the first decay in the density profile.
After surface height becomes comparable to the characteristic length of a polymer
for a given $G$ value, such that there are several structures of polymer lumps,
the incident monomer with a certain angle can hang on the other polymer bodies
and both the effect of the cosine flux and the diffusion of monomers
governs the growing dynamics; this reflects the plateau in the density profile.
Finally, the front of surface is governed by the fluctuations of the locations of active ends,
which is shown as the second decay in the density profile. We wish to note here that the
density profile is taken at $t=100$ ML, which corresponds to the same as the right side panels
in figure~\ref{snapshots}.

Although the qualitative behaviors of kinetic roughening seem to
be similar for all three cases of $G$, its quantitative behavior
quite depends on the value of $G$. Such a role of diffusion in the
VDP growth is summarized as the $G$-dependent kinetic roughening
in table~\ref{table}, in terms of the roughness exponent, $\alpha$,
and the dynamic exponent, $z$, for both the global and local cases.
It should be noted that the growth exponent $\beta$ that
we found above is different from either $\alpha_{\rm global}/z_{\rm global}$
or $\alpha_{\rm local}/z_{\rm local}$.
Therefore, the data of $W$ hardly collapse
due to the VDP growth regime (see figure~\ref{G10W-collapse}).
\begin{table}[t]
\caption{Summary of roughness exponents and dynamic exponents for various $G$:
Unlike the global results, the local results seems to be independent
of the $G$ value. \label{table} }
\begin{center}
\begin{tabular}{@{\extracolsep{\fill}} l | r r r r }
\hline\hline
 G & $\alpha_{\rm global}$ & $\alpha_{\rm local}$ & $z_{\rm global}$ & $z_{\rm local}$ \\
 \hline
 $10$ & 0.89(1) & 0.50(2) & 1.27(1) & 1.27(2) \\
 $10^3$  & 0.87(1) & 0.47(2) & 1.16(1) & 1.27(2) \\
 $10^5$  & 0.72(1) & 0.48(2) & 0.81(1) & 1.32(2) \\
\hline\hline
\end{tabular}
\end{center}
\end{table}

\subsection{Height and step distributions}

\begin{figure}[t]
\begin{center}
\includegraphics[width=0.35\columnwidth]{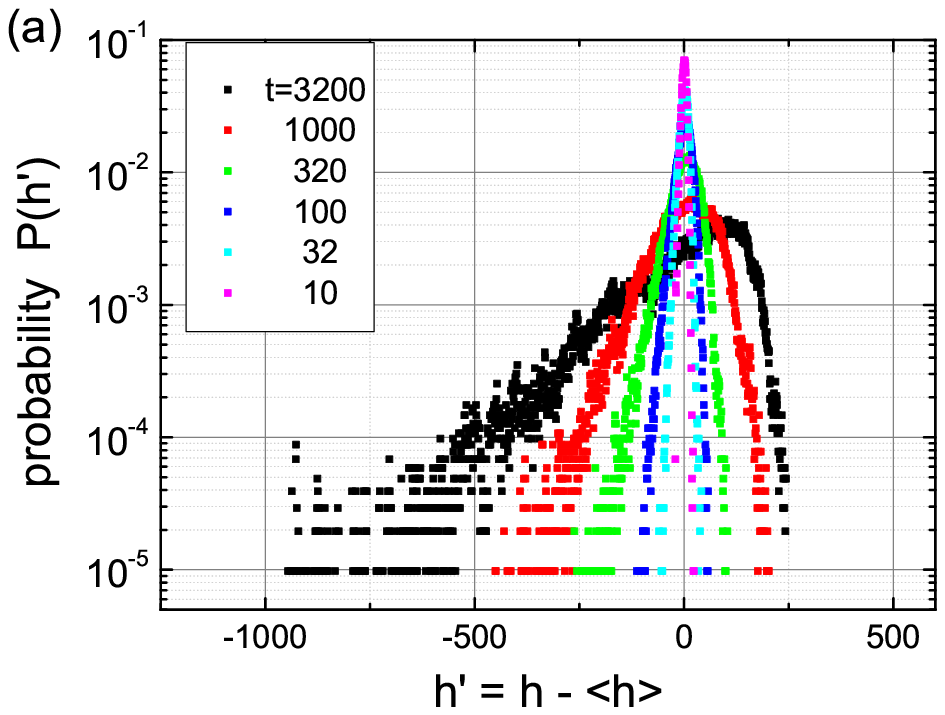}~~~~~
\includegraphics[width=0.35\columnwidth]{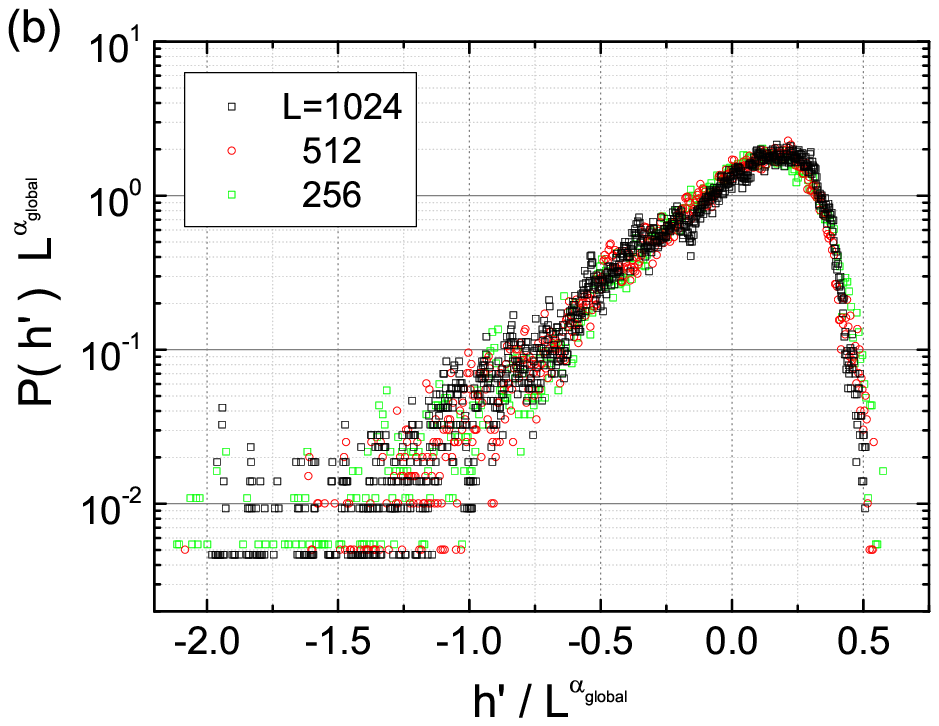}
\end{center}
\caption{(a) Semi-logarithmic plots of the height distribution function, $P(h')$,
against $h'=h-\langle{h}\rangle$, at various times, $t=10,~32~,100,~320,~1000$,
and $3200$ ML for $L=1024$ and $G=10$. Note that $\langle h\rangle=\bar{h}$ for our case.
As $t$ elapses, $P(h')$ becomes broader and is gradually transformed into a right skewed Gaussian distribution.
(b) At $t=3200$ ML, after the surface roughness gets saturated,
$P(h')$ exhibits scaling behavior with $\alpha_{\rm global}$, which is confirmed
for various system sizes $L=256,~512$, and $1024$,
as $P(h')L^{\alpha_{\rm global}}$ versus $h'/L^{\alpha_{\rm global}}$
with $\alpha_{\rm global}=0.89$ for $G=10$.
 \label{Ph}}
\end{figure}

In measuring the height distribution, $P(h')$ where $h'=h-\langle h\rangle$,
for various times and system sizes, we double-check the anomalous kinetic roughening
in our VDP growth model and also confirm our numerical finding of $\alpha_{\rm global}$
by collapsing the data (see figure~\ref{Ph}) of $P(h')$.
The height distribution becomes broader as time elapses,
which means the surface gets rougher for the larger value of the width $W$
since $W$ corresponds to the standard deviation of $P(h')$. At the initial stage,
$P(h')$ is almost Gaussian and symmetric, while at the final stage,
the distribution is slightly skewed to the right, where the exponential decay tail
below side of the average height (left) is broader
than that above it (right).

\begin{figure*}[]
\begin{center}
\includegraphics[width=0.35\columnwidth]{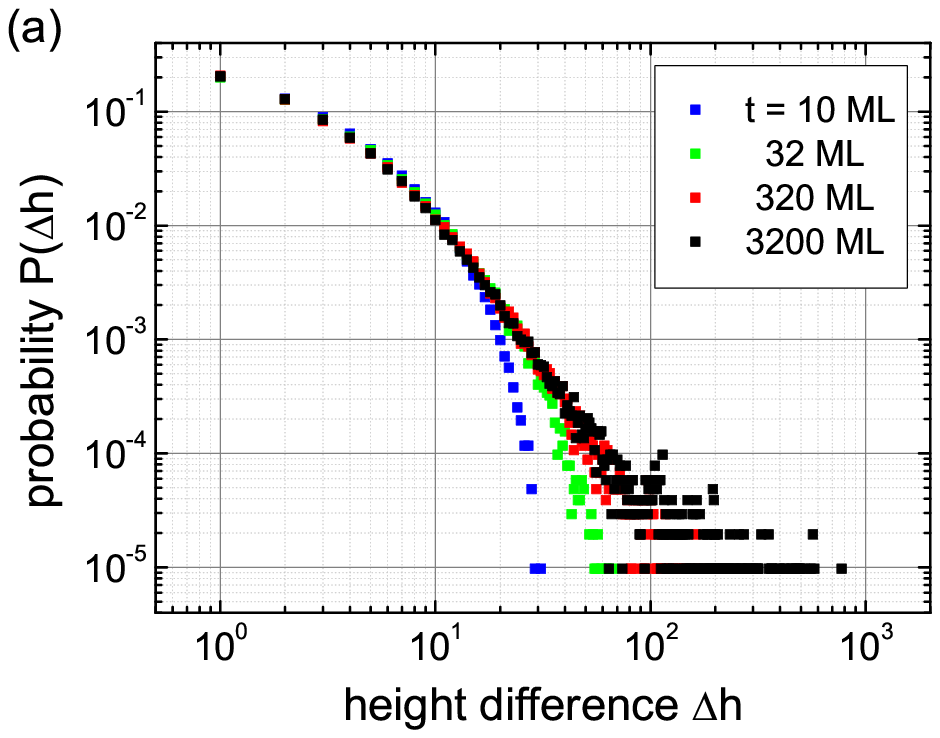}~~~~~
\includegraphics[width=0.375\columnwidth]{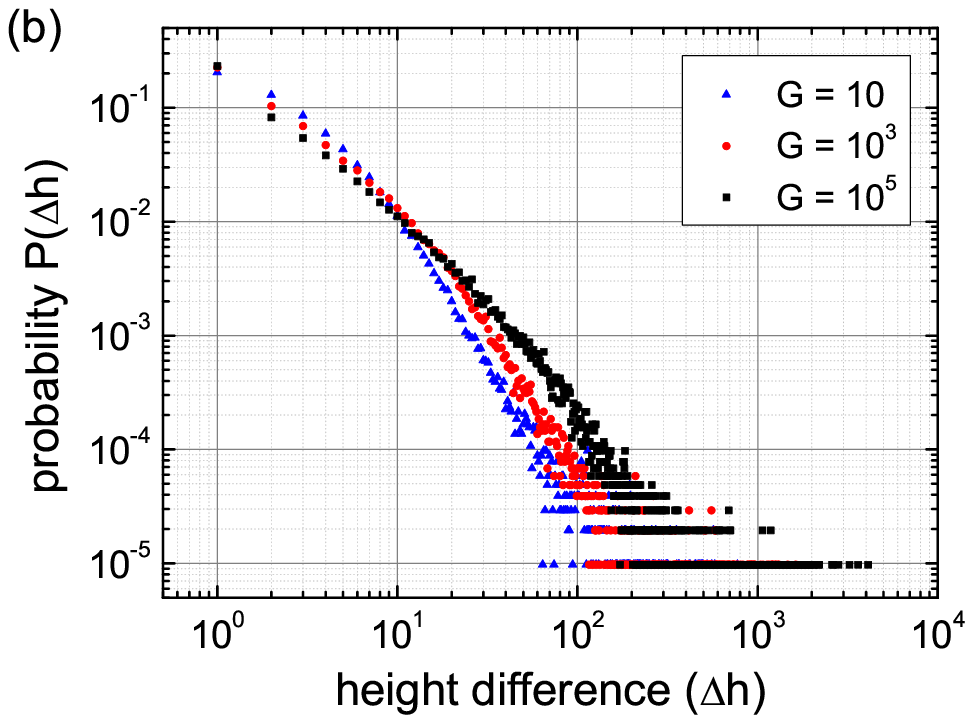}
\end{center}
\caption{Power-law step distributions for $L=1024$:
(a) at various times only for $G=10$ and
(b) for three values of $G$ only at $t=3200$ ML. \label{Pdh}}
\end{figure*}

It is observed that anomalous kinetic roughening in the VDP growth
is attributed to the power-law distribution of the height difference
among the nearest-neighboring sites, i.e., $r=1$ (namely `step'), for
$P(\Delta h)$, which implies that the VDP growth exhibits multifractality
as well as $\alpha_{\rm global}\ne\alpha_{\rm local}$.
We investigate how the power-law behavior of $P(\Delta h)$ changes as $t$ elapses
and as $G$ increases.  Figure~\ref{Pdh} shows that
for the larger values of $\Delta h$ the decay exponent seems to be independent of the $G$ value
in the stead-state limit. It is very interesting
that the step distribution shows clearly a power-law decay
for large values $\Delta h(\equiv |h(x+1)-h(x)|)$ after $W$ gets saturated.

This is somewhat similar to the case for the ballistic deposition model
with a power-law noise~\cite{BD+power-law}.  In that sense,
we suspect that the active ends play a crucial role
in the power-law step distribution, the details of which
are under investigation~\cite{ongoing-VDP1D}.

\subsection{Polymer Properties}
\begin{figure*}[t]
\begin{center}
\includegraphics[width=0.30\columnwidth]{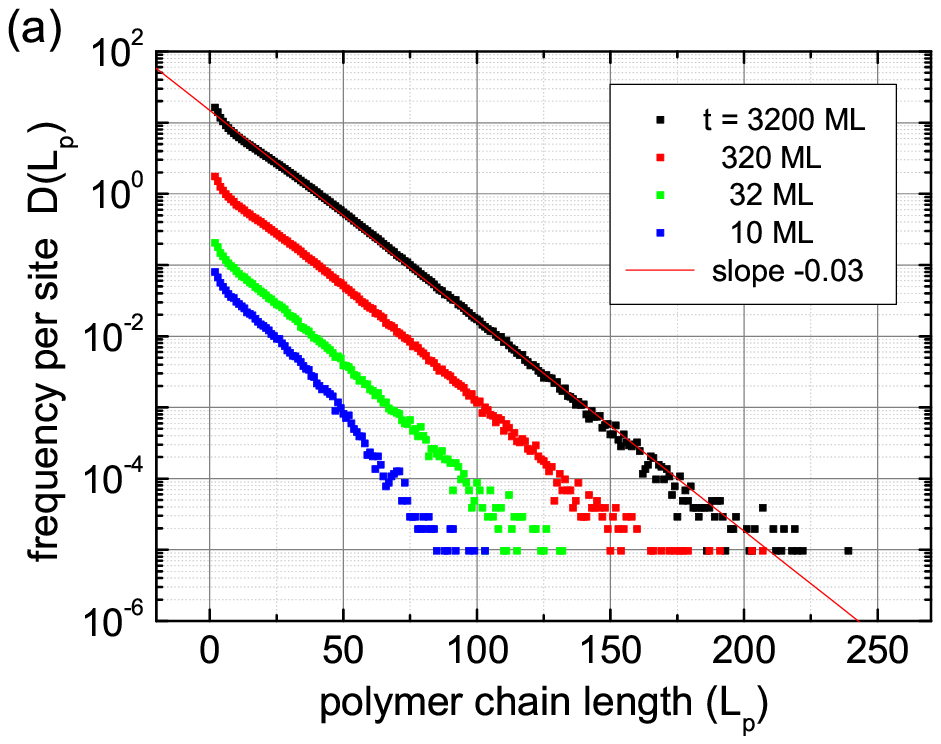}\hfill
\includegraphics[width=0.315\columnwidth]{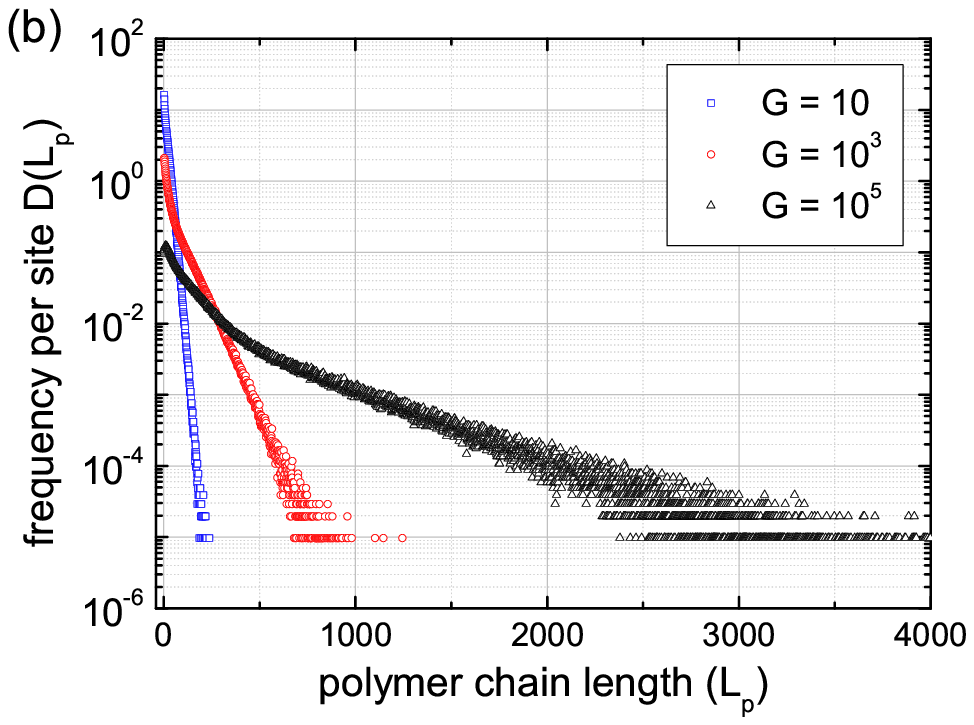}\hfill
\includegraphics[width=0.30\columnwidth]{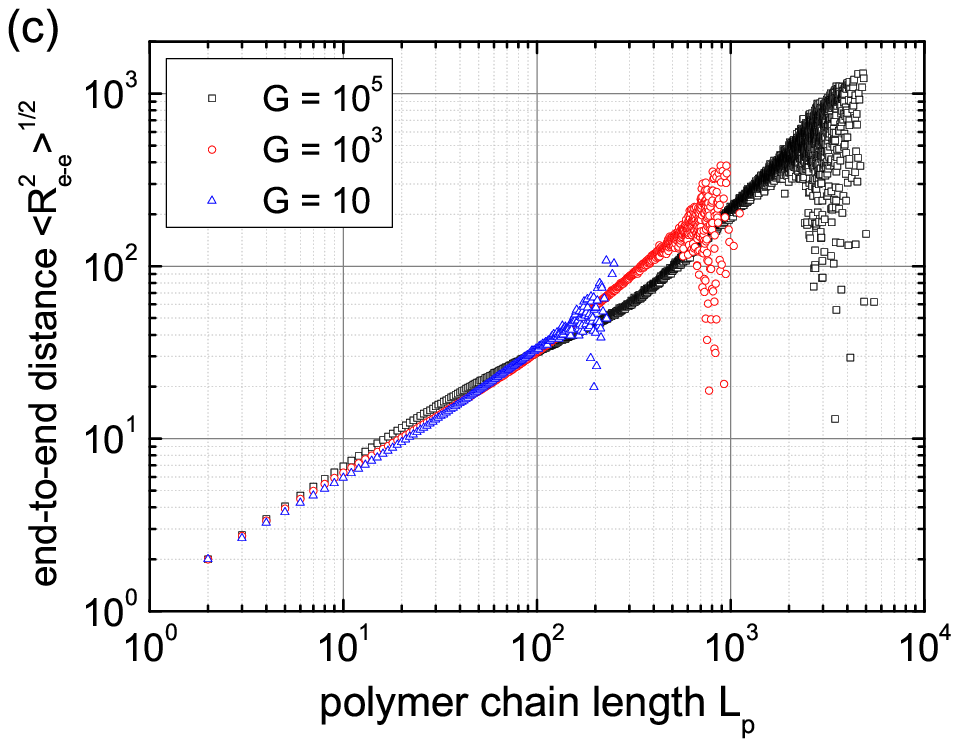}
\end{center}
\caption{Semi-logarithmic plots of the polymer
chain length distribution $D(L_p)$ against the length of polymer
$L_p$ (a) for $G=10$ at various times and (b) for various $G$ values at $t=3200$ ML,
where $L=1024$. For the same setup as (b), (c) shows double-logarithmic plots of the
end-to-end distance of polymer, $\langle R_{e-e}^2 \rangle^{1/2}$, against $L_p$.
\label{polymer}}
\end{figure*}

In the VDP growth, the properties of the polymer are also important
(to be discussed). After measuring the time-dependent frequency
of the polymers per each site, $D(L_p)$, where $L_p$ is the length
of polymer for $L=1024$, as well as the end-to-end distances (see
figure~\ref{polymer}), we finally investigate such properties.
As time elapses, monomers are deposited on the surface more and more,
so the number of polymers increases and at the same time
polymers get longer. On the basis of our numerical finding,
there is a typical length scale of polymers for a given value of $G$
in the steady state of the VDP growth. It is observed that
the typical length of a polymer gets longer as $G$ increases
(see figure~\ref{polymer} (b)). For example, one typical polymer consists of
about 15 monomers at $G=10$, while 434 monomers at $G=10^5$.
Figure~\ref{polymer} (c) shows that the root-mean-square of the end-to-end distance
for a given polymer, $\langle R_{\rm e-e}^2 \rangle^{1/2}$, scales as
$\langle R_{\rm e-e}^2 \rangle^{1/2} \propto L_p^\nu$,
where we find that the exponent $\nu$ is about 0.75 for short polymers
under about 100 monomer length, but 1.0 for long polymers.
Therefore, as $G$ increases, it is observed that a crossover
from $\nu=0.75$ to $\nu=1$ occurs. Here, the exponent $\nu$ represents
the inverse of the fractal dimension of polymers.
One can say $D_{\rm f}=1.33$ at $G=10$, which is the same as that of
the linear polymers formed by self-avoiding walks~\cite{Meakin1998B}.
The detailed analysis has been investigated~\cite{ongoing-VDP1D}.

\subsection{Growth of (2+1)-dimensional VDP thin films}

The (2+1)-dimensional version of our model has been also considered
in order to explain the most recent experimental results
by Lee and his co-workers~\cite{IJLee2008};
the growth of PPX-C films was discussed by the same authors. 
It is noted that our extended version can be considered
as a modification of the earlier study
by Zhao and his co-workers~\cite{2D-submonolayer-VDP}
for the VDP process in the submonolayer regime.
In our extension, multilayer growth is allowed,
with the coalescence process of polymers.
Our preliminary results in a (2+1)-dimensional lattice~\cite{ongoing-VDP2D}
seems to be quite different from that in a (1+1)-dimensional lattice,
but they also exhibit anomalous scaling behavior in kinetic roughening
with multifractality; this is similar to the experimental results,
except that the valley filling regime seems to be missing in our model study.
To answer the question of the origin of the valley filling regime of the experimental results,
it might be necessary that we also consider some new dynamics,
such as chain relaxations we ignored in our current version.
Considering polymer properties in the VDP growth would be another key
to identifying the universality class of the VDP growth more clearly.
For example, we suspect that the reptation with zigzag paths
governs the sublinear scaling behavior at early stage of the polymer growth,
while the polymer interaction become relevant after polymers grow
enough to be comparable with the typical length,
so the coalescence of polymers let them show the linear scaling,
as shown in the (1+1)-dimensional version. Such properties have been also investigated
in our modified version in a (2+1)-dimensional lattice~\cite{ongoing-VDP2D}.

\section{Summary and remarks}
\label{sec:conclusion}

In summary, we studied a simple toy model for the growth of polymer thin films by vapor deposition polymerization (VDP) processes in order to explain recent experimental results for the coating processes of poly (p-xylylene) (PPX) and the derivatives, e.g., PPX-C. It is found that the VDP growth is quite different from the conventional molecular beam expitaxy (MBE) growth for the growth of metal or semiconductor films. In particular, we argued that anomalous scaling behavior in kinetic roughening for the VDP growth is attributable to the instability induced by the non-local shadowing effects as well as active bonding in polymerization. As another clear evidence of such anomalies, we showed the power-law step distributions, directly related to the multifractality of the VDP growth. The two-point height difference $q$th-moment analyses are also under detailed investigation~\cite{ongoing-VDP1D,ongoing-VDP2D}.

Finally, we would like to comment on polymer interaction, i.e., the coalescence process of polymers, which is the new aspect of our model. In earlier studies, polymer interaction and chain relaxation are often omitted from dynamic rules due to their complexity in model simulation codes. We retained polymer interactions since they play a crucial role in the comparison of real experimental data, in particular for the polymer structure and its growth, while we also omitted the chain relaxation rule due to the same reason.
In the model without polymer interaction studied by Bowie and Zhao~\cite{Bowie2004}, the number of active ends of polymers always increases, since there is no mechanism for reducing the number of polymers, while in our model, the increment of polymers slows down as polymers are merged into others. One polymer interaction removes two active ends, and the active ends are the stabilizing sites of monomers in the VDP growth model. Thus, one can readily anticipate that the monomers in our model is more abundant as compared to the case for excluding the coalescence process. Moreover, the diffusion probability $P_D=\frac{G\rho_m}{G\rho_m+1}$ can effectively increases when the ratio of diffusion rate to deposition flux rate $G$ is compatible with the monomer density $\rho_m$. Of course, such an effect becomes negligible when $G\gg\rho_m$ since the number of polymers is small, so the polymer coalescence process rarely happens.
Regarding the effects of the coalescence of polymers on the polymer structure and its growth in our model,
we have observed that in the characteristic polymer length definitely becomes longer and the fractal dimension of polymers gets clearly larger from $D_f=1.08$ to $D_f=1.33$ (closer to that of self-avoiding walk polymers), as compared to excluding polymer interaction~\cite{Bowie2004,ongoing-VDP1D}. In contrast to the dramatic polymer structural change, surface roughnesses behave in almost the same way in two cases even though the number of monomers in our model rapidly increases as compared to the case when excluding polymer interaction, as expected. Therefore, we conclude that the polymer interaction mechanism gives us a better understanding of the polymer structural properties than its growth properties in the (1+1)-dimensional case. The role of such a mechanism in the (2+1)-dimensional case will be discussed elsewhere~\cite{ongoing-VDP2D}.
%
%
\ack{This work was supported by the BK21 project (MH)
and Acceleration Research (CNRC) of MOST/KOSEF
through the grant No. R17-2007-073-01001-0 (SS)
and by the Korean Systems Biology Program from MEST
through KOSEF (No. M10309020000-03B5002-00000, HJ).
We would like to acknowledge fruitful discussions with I.J. Lee,
who gave us the main idea of this work and let it being initiated,
and valuable comments from J. Krug and J.M. Kim.}


\section*{References}
\begin{thebibliography}{10}

\bibitem{FnV1991B+BnS1995B}
{\it Dynamics of Fractal Surfaces}, edited by Family F and Vicsek T
(World Scientific, Singapore, 1991);
Barab{\'a}si A-L and Stanley H E,
1995 {\it Fractal Concepts in Surface Growth}
(Cambridge: Cambridge University Press)

\bibitem{Meakin1998B}
Meakin P, 1998 {\it Fractals, Scaling and Growth far from Equilibrium}
(Cambridge: Cambridge University Press)

\bibitem{Wong1993B}
{\it Polymers for Electronic and Photonic Applicants},
edited by Wong C P (Academic Press, Boston, 1993)

\bibitem{Collins1994+Biscarini1997}
Collins G W, Letts S A, Fearon E M, McEachern R L, and Bernat T P,
1994 Phys. Rev. Lett. {\bf 73}, 708;
Biscarini F, Samor{\'i} P, Greco O, and Zamboni R,
1997 Phys. Rev. Lett. {\bf 78}, 2389

\bibitem{Zhao2000}
Zhao Y-P, Fortin J B, Bonvallet G, Wang G-C, and Lu T-M,
2000 Phys. Rev. Lett. {\bf 85}. 3229;
Punyindu P and Das Sarma S, 2001 Phys. Rev. Lett. {\bf 86}, 2696;
Zhao Y-P, Fortin J B, Bonvallet G, Wang G-C, and Lu T-M,
2001 Phys. Rev. Lett. {\bf 86}. 2697

\bibitem{IJLee2008}
Lee I J, Yun M, Lee S-M, and Kim J-Y, 2008 Phys. Rev. B {\bf 78}, 115427

\bibitem{VDP-PPX}
Beach W F, 1977 Macromolecules {\bf 11}, 72;
Beach W F, Lee C, Basset D R, Austin T M, and Olson O, 1989
in {\it Encyclopedia of Polymer Science and Engineering}
(Wiley, New York, 1989), 2nd ed., Vol. 7, p990

\bibitem{Bowie2004}
Bowie W and Zhao Y-P, 2004 Surf. Sci. {\bf 563}, L245;
Zhao Y-P and Bowie W, 2005 Mater. Res. Soc. Symp. Proc. {\bf 859E}, JJ6.4.1

\bibitem{cosineF}
Drotar J T, Zhao Y-P, Lu T-M, and Wang G-C,
2000 Phys. Rev. B. {\bf 62}, 2118;
Karabacak T, Zhao Y-P, Wang G-C, and Lu T-M,
2001 Phys. Rev. B {\bf 64}, 085323;
Yang Y G, Hass D D and Wadley H N G, 2004 Thin Solid Films, {\bf 471}, 1;
Yanguas-Gil A, Cotrino J, Barranco A, and Gonz{\'a}lez-Elipe A R,
2006 Phys. Rev. Lett. {\bf 96}, 236101;
Pelliccione M, Karabacak T, and Lu T-M,
2006 Phys. Rev. Lett. {\bf 96}, 146105;
Pelliccione M, Karabacak T, Gaire C, Wang G-C, and Lu T-M,
2006 Phys. Rev. B {\bf 74}, 125420

\bibitem{BD+power-law}
Zhang Y-C, 1990 Physica A {\bf 170} 1;
Krug J, 1991 J. Phys. I France {\bf 1}, 9;
Buldyrev S V, Havlin S, Kertesz J, Stanley H E, and Vicsek T,
1991 Phys. Rev. A {\bf 43}, 7113;
Lam C-H and Sander L M, 1992 Phys. Rev. Lett. {\bf 69}, 3338;
Lam C-H and Sander L M, 1992 J. Phys. A: Math. Gen. {\bf 25}, L135;
Lam C-H and Sander L M, 1993 Phys. Rev. E {\bf 48}, 979;
Barab{\'a}si A-L, Bourbonnais R, Jensen M, Kertesz J, Vicsek T, and Zhang Y-C,
1992 Phys. Rev. A {\bf 45}, R6951

\bibitem{ongoing-VDP1D}
Son S-W, Ha M, Jeong H, and Lee I J, 2009 in preparation

\bibitem{2D-submonolayer-VDP}
Zhao Y-P, Hopper A R, Wang G-C, and Lu T M,
1999 Phys. Rev. E {\bf 60}, 4310 (1999);
Zhao Y-P, Hopper A R, Wang G-C, and Lu T M,
2000 Phys. Rev. E {\bf 61}, 2156

\bibitem{ongoing-VDP2D}
Son S-W, Ha M, Lee I J, and Jeong H, 2008 unpublished data

\end{thebibliography}
\end{document}